\documentclass[preprint,12pt]{elsarticle}

\usepackage{graphicx}
\graphicspath{{Figures/}}
\usepackage{amsmath,amsfonts,amssymb}

\usepackage{subfigure}

\usepackage{color}

\journal{Computers and Fluids}

\newcommand\blfootnote[1]{%
  \begingroup
  \renewcommand\thefootnote{}\footnote{#1}%
  \addtocounter{footnote}{-1}%
  \endgroup
}

\usepackage{amsthm}
\usepackage{setspace}
\newtheorem{remark}{Remark}

\begin{document}

\begin{frontmatter}


\title{{A turbulent eddy-viscosity surrogate modeling framework for Reynolds-Averaged Navier-Stokes simulations}}




\author[label1]{Romit Maulik}
\ead{rmaulik@anl.gov}
\address[label1]{Argonne Leadership Computing Facility, Argonne National Laboratory, Lemont, IL-60439, USA}

\author[label1]{Himanshu Sharma}

\author[label2]{Saumil Patel}
\address[label2]{Computational Science Division, Argonne National Laboratory, Lemont, IL-60439, USA}

\author[label1]{Bethany Lusch}
\author[label1,label3]{Elise Jennings}
\address[label3]{Irish Centre for High-End Computing (ICHEC), The Tower, Trinity Technology \& Enterprise Campus, Grand Canal Dock, Dublin 2, Ireland}

\begin{abstract}
The Reynolds-averaged Navier-Stokes (RANS) equations for steady-state assessment of incompressible turbulent flows remain the workhorse for practical computational fluid dynamics (CFD) applications. Consequently, improvements in speed or accuracy have the potential to affect a diverse range of applications. We introduce a machine learning framework for the {surrogate modeling of steady-state turbulent eddy viscosities for RANS simulations, given the initial conditions. This modeling strategy} is assessed for parametric interpolation, while numerically solving for the pressure and velocity equations to steady state, thus representing a framework that is hybridized with machine learning. We achieve {competitive} steady-state results with a significant reduction in solution time when compared to those obtained by the Spalart-Allmaras one-equation model. This is because the proposed methodology allows for considerably larger relaxation factors for the steady-state velocity and pressure solvers. Our assessments are made for a backward-facing step with considerable mesh anisotropy and separation to represent a practical CFD application. For test experiments with \textcolor{black}{either} varying inlet velocity conditions or step heights we see time-to-solution reductions around a factor of 5. The results represent an opportunity for the rapid exploration of parameter spaces that prove prohibitive when utilizing turbulence closure models with multiple coupled partial differential equations. \blfootnote{Code available publicly at \texttt{https://github.com/argonne-lcf/TensorFlowFoam}}
\end{abstract}


\end{frontmatter}


\section{Introduction}

The direct numerical simulation (DNS) of turbulence is impractical for most applications due to its multiscale nature. Engineering decisions requiring a short turnover duration continue to rely on the Reynolds-averaged Navier-Stokes (RANS) equations for steady-state analysis of turbulent flows. Within that context, developments that lead to improved RANS results from the point of view of speed or accuracy have the potential to affect design workflows significantly. To that end, there have been several recent investigations into the augmentation of RANS using data-driven methods that attempt to improve its accuracy through the use of fidelity porting strategies \cite{zhang2015machine,ling2015evaluation,xiao2016quantifying,weatheritt2016novel,ling2016reynolds,wu2018physics,wang2017physics,sotgiu2019towards,wu2019physics,cruz2019use,geneva2019quantifying,layton2020diagnostics} or the utilization of experimental data \cite{parish2016paradigm,singh2017machine}. Closure optimization strategies have also looked at zonal coefficient refinement or model selection for improved accuracy \cite{matai2019zonal,maulik2019sub}. Most attempts at infusing information from higher fidelity models such as DNS to reduced-order representations have resulted in improved accuracy for the quantities of interest and represent the bulk of physics-informed machine learning investigations of turbulence closure modeling \cite{gamahara2017searching,maulik2017neural,vollant2017subgrid,maulik2018data,beck2019deep}. Recently, researchers have also advocated for CFD-driven machine learning wherein epochs of the machine learning optimization are coupled with a forward solve of the RANS equations \cite{zhao2019turbulence,taghizadeh2020turbulence} and have also investigated ideas of spatial and spatio-temporal super-resolution for reconstructing sub-grid information \cite{fukami2019super,fukami2020machine,maulik2020probabilistic}. The numerous opportunities for utilizing machine learning in turbulence modeling have also been summarized in recent reviews such as \cite{duraisamy2019turbulence} and \cite{zhang2019recent}. 

We are motivated by the seminal work of Tracey, Duraisamy and Alonso \cite{tracey2015machine}. This study assessed the feasibility of using neural networks to predict turbulent eddy viscosities obtained from training data generated by the Spalart-Allmaras model. We also note a recent investigation by Zhu et al., \cite{zhu2019machine} which looked at the surrogate modeling of subsonic flow around an airfoil with a radial-basis function network. The unique feature of this study was a section devoted to the improvements in compute time due to the preclusion of a separate partial differential equation (PDE) for the turbulent eddy viscosity computation. In contrast to the above investigations, this article investigates an alternate {approach for eddy viscosity surrogate modeling through the \emph{direct prediction} of steady-state turbulent eddy viscosity fields with the use of a deep learning framework}. We leverage the initial conditions of our test case (given by a low fidelity solution such as potential flow) across a range of boundary conditions as inputs and harvest steady-state turbulent eddy viscosities (from a suitable choice of a turbulence closure model based on the linear eddy-viscosity hypothesis) as outputs to train a relationship which need to be deployed \emph{once} at the start of a RANS simulation. In addition, we may also incorporate direct knowledge of the control parameter to improve the accuracy of this surrogate map. The velocity and pressure equations are solved to steady-state while utilizing this \emph{fixed} solution. It is observed that the resultant RANS solutions are able to approximately replicate the trends of the solutions of the PDE model but with the added advantage of significant reductions to the total time-to-solution. \textcolor{black}{The aforementioned study is also performed using training data for different geometries to test the viability of the framework for predicting steady-state turbulent eddy-viscosities in case of varying parameters that control the computational mesh.}

The machine learning (ML) framework presented here consists of two phases, which we refer to as {\it offline} and {\it online}. The offline phase consists of data generation at strategically selected points in parameter space (corresponding to a design of experiments component) as well as the training of a neural network to predict a turbulent eddy-viscosity. This trained network is the ML surrogate model. The online phase corresponds to the deployment of the trained ML surrogate to predict a spatial field of eddy-viscosities which are then held fixed, while the velocity and pressure equations are solved to convergence. The overall procedure with online and offline phases for training and deploying the ML surrogate is denoted ``the ML framework''. Our test case is given by the two-dimensional backward-facing step exhibiting flow separation and considerable mesh anisotropy and thus represents a challenging test for the construction of surrogate closures for real-world applications. We note that the mesh utilized in this study, which will be introduced in greater detail in Section \ref{Problem}, is generated with information from the NASA LARC extended turbulence validation database.

\section{Methodology}

This section describes the proposed data-driven workflow and test case. In subsections \ref{Problem} and \ref{subsection:RANS} we define the problem statement and provide a brief background of RANS modeling. In subsection \ref{subsection:ml-rans} a description of the proposed ML and computational fluid dynamics (ML-CFD) integration procedure is provided.

\subsection{Problem Definition}\label{Problem}
\begin{figure}
    \centering
    \includegraphics[scale=0.25,width=\textwidth]{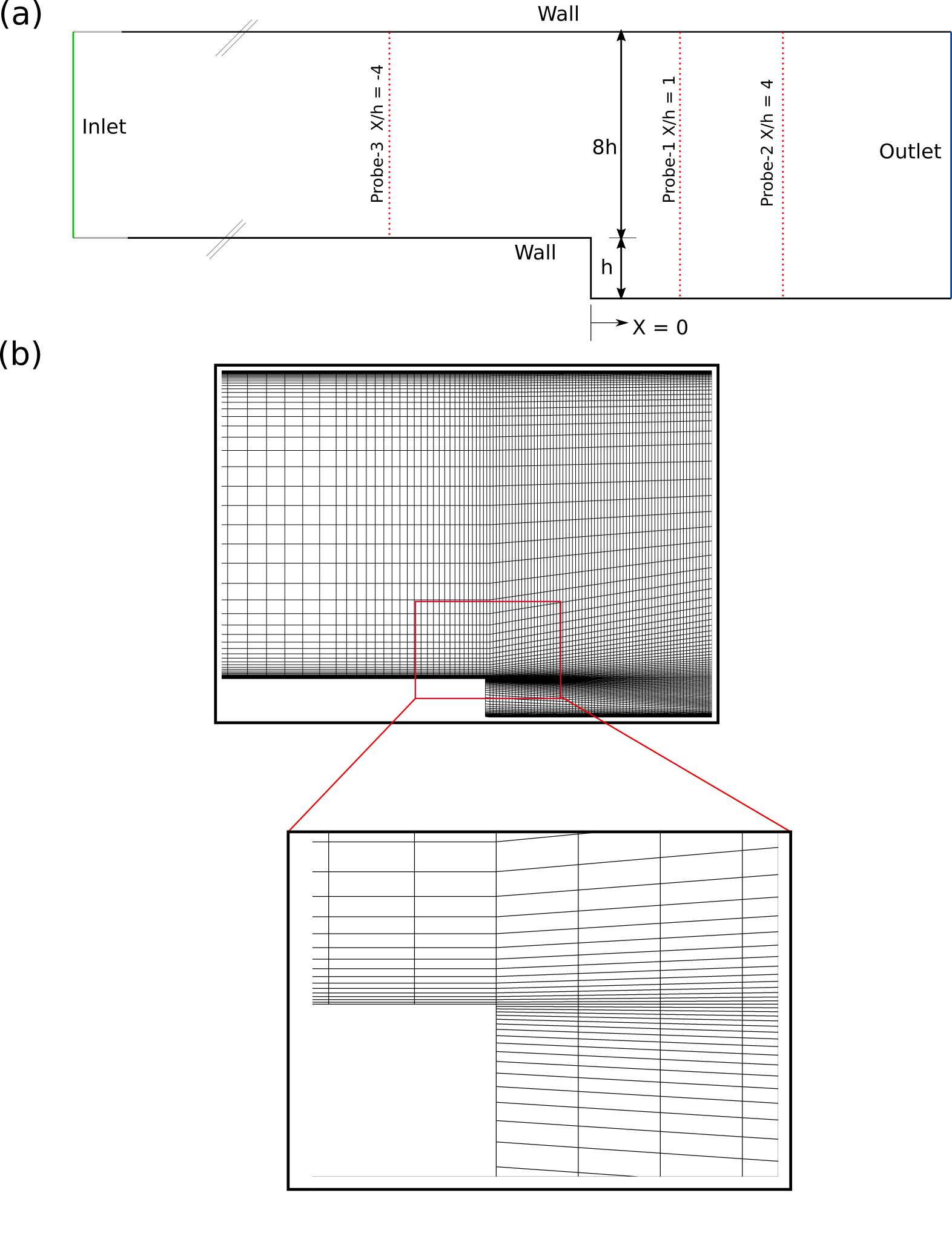}
    \caption{The backward step used for a representative CFD simulation in this investigation. (a) shows the geometry and the step defined at (0,0). Red dashed lines show locations where data has been probed to assess our ML surrogate model. (b) shows the mesh used for the computation and a zoomed view of the refined mesh.}
    \label{fig:mesh_Geo}
\end{figure}

In the current work, we undertake the problem of simulating flow past the backward-facing step as shown in figure \ref{fig:mesh_Geo}(a). The experimental description is provided in \cite{driver1985features}. For our first experiment, we simulate the airflow for a Reynolds number based on the fixed step height of  $h = 1.27$ cm. The Reynolds number is defined as $Re_{h}= \frac{U h}{\nu}$ where $U$ is the freestream velocity (and also a control parameter for generating training and testing data in our machine learning problem definition) and $\nu = 1.5 \times 10^{-5} \frac{m^2}{s}$ is the kinematic viscosity. The problem is widely used for the validation of novel turbulence models. More details about the case can be seen on the NASA turbulence web page\footnote{https://turbmodels.larc.nasa.gov/backstep\_val.html}.
\\
\\
\indent The training data set for the first experiment is generated from ten simulations, each with distinct free-stream velocities resulting in Reynolds numbers that range from $Re_h = 34,000 \text{-} 41,500$. The computational domain is composed of a non-uniform grid as shown in figure \ref{fig:mesh_Geo}. To capture the separation occurring near the step, additional refinement is performed on the grid, which can be seen in the zoomed view shown in Figure \ref{fig:mesh_Geo}(b). The total number of cells used in this experiment was fixed at $20,540$. The mesh was generated using OpenFOAM's native grid generator \texttt{blockMesh}, with a maximum aspect ratio of $7,601.12$. 

A second experiment is designed by utilizing a fixed inlet velocity $U_b=44.2 m/s$ while varying the step height. We select $0.50h$, $0.75h$, $1.25h$, $1.5h$, $2.0h$ as our different data generation geometries and re-mesh the entire domain for each option.
Table \ref{TableCell} provides a summary of the various mesh configurations.
\begin{table}
\centering
\begin{tabular} {l*{2}{c}}
Step height & Cell count \\
\hline
0.5h  &  130,184  \\
0.75h &  134,648  \\
1.25h &  128,808  \\
1.50h &  130,824  \\
1.75h &  152,470  \\
1.90h &  131,374  \\
2.00h &  131,948  \\
\end{tabular}
\caption{Mesh configurations showing different step heights and associated cell count} 
\label{TableCell}
\end{table}
This experiment would thus correspond to a numerical simulation campaign that scans across a family of geometries with the ML surrogate for $\nu_t$ aiding in faster times to solution for each configuration. All CFD simulations are performed with the OpenFOAM solver (version 5.0) \cite{weller1998tensorial} which has seen extensive use in practical CFD applications \cite{robertson2015validation}

\subsection{RANS}\label{subsection:RANS}

The RANS equations are a time-averaged form of the Navier-Stokes equations of motion for fluid flow. The equations are derived using the Reynolds decomposition principle shown in \cite{reynolds1895iv}, whereby the instantaneous quantity is decomposed into its time-averaged and fluctuating quantities. These equations are used for describing the steady-state behavior of turbulent flows. In Cartesian coordinates for a stationary flow of an incompressible Newtonian fluid, the RANS equations can be written as follows:
\begin{subequations}
\begin{equation}
    \frac{\partial  \bar{u_i} }{\partial x_i}= 0,
\end{equation}
\begin{equation}
\centering
\rho \bar{u_j}\frac{\partial  \bar{u_i} }{\partial x_j} = \rho \bar{f_i} + \frac{\partial}{\partial x_j} 
\left[ - \bar{p}\delta_{ij} + 2\mu \bar{S_{ij}} - \rho \overline{u_i^\prime u_j^\prime} \right]
\label{Eq:RANS}
\end{equation}
\end{subequations}
where $\bar{.}$ denotes time-averaged quantities, ${u}$ is the velocity component, $u^\prime$ is the fluctuating component, $p$ is the pressure, $\rho$ is the density of fluid, $f_{i}$ is a vector representing external forces, $\delta_{ij}$ is the Kronecker delta function, $\mu$ is the dynamic viscosity, and  $\bar{S_{ij}} = \frac{1}{2}\left( \frac{\partial \bar{u_i}}{\partial x_j} + \frac{\partial \bar{u_j}}{\partial x_i} \right)$ is the mean rate of the stress tensor. The left-hand side of this equation defines the change of the mean momentum of the fluid element due to the unsteadiness of the mean flow and the convection. The change is balanced by the mean body force, the isotropic stress resulting from the mean pressure field, the viscous stresses and the stresses due to the fluctuating velocity field ${\displaystyle \left(-\rho {\overline {u_{i}^{\prime }u_{j}^{\prime }}}\right)}$, generally denoted the \textit{Reynolds stress}. This nonlinear term requires an additional model specification to close the RANS equation. Most methods deal with an explicit model for this tensor through the utilization of additional algebraic or differential equations. Some of these are based on the time evolution of the Reynolds stress equation as presented in \cite{chou1945velocity}. The bulk of our experiments in this investigation utilizes the one-equation Spalart-Allmaras \cite{spalart1992one} model (SA) for the generation of reference data sets. \\
\indent The SA model closure is written in the following form:
\begin{equation}
\begin{aligned}
{\frac {\partial {\tilde {\nu }}}{\partial t}}+u_{j}{\frac {\partial {\tilde {\nu }}}{\partial x_{j}}} &= C_{b1}[1-f_{t2}]{\tilde {S}}{\tilde {\nu }}+{\frac {1}{\sigma }}\{\nabla \cdot [(\nu +{\tilde {\nu }}) \nabla {\tilde {\nu }}]+C_{b2}|\nabla {\tilde {\nu }}|^{2}\}\\
&-\left[C_{w1}f_{w}-{\frac {C_{b1}}{\kappa ^{2}}}f_{t2}\right]\left({\frac {\tilde {\nu }}{d}}\right)^{2}+f_{t1}\Delta U^{2}
\end{aligned}
\end{equation}
\begin{align*}
&\nu _{t}={\tilde  {\nu }}f_{{v1}}, f_{{v1}}={\frac  {\chi ^{3}}{\chi ^{3}+C_{{v1}}^{3}}}, \chi :={\frac  {{\tilde  {\nu }}}{\nu }},{\tilde {S}}\equiv S+{\frac  {{\tilde  {\nu }}}{\kappa ^{2}d^{2}}}f_{{v2}}, f_{v2} = 1 -{\frac{\chi }{1+\chi f\_{{v1}}}},\\
&f_{w}=g\left[{\frac{1+C_{{w3}}^{6}}{g^{6}+C_{{w3}}^{6}}}\right]^{{1/6}},g=r+C_{{w2}}(r^{6}-r), r\equiv {\frac  {{\tilde  {\nu }}}{{\tilde{S}}\kappa ^{2}d^{2}}},S={\sqrt  {2\Omega _{{ij}}\Omega _{{ij}}}},\\ 
& f_{{t1}}=C_{{t1}}g_{t}\exp \left(-C_{{t2}}{\frac  {\omega _{t}^{2}}{\Delta U^{2}}}[d^{2}+g_{t}^{2}d_{t}^{2}]\right),
f_{{t2}}=C_{{t3}}\exp \left(-C_{{t4}}\chi ^{2}\right),\\
& \Omega _{{ij}}={\frac  {1}{2}}(\partial u_{i}/\partial x_{j}-\partial u_{j}/\partial x_{i}),\sigma=2/3,C_{{b1}}=0.1355,C_{{b2}}=0.622,\kappa=0.41,\\
& C_{{w1}}=C_{{b1}}/\kappa^{2}+(1+C_{{b2}})/\sigma, C_{{w2}}=0.3,C_{{w3}}=2,C_{{v1}}=7.1,\\
& C_{{t1}}=1,C_{{t2}}=2,C_{{t3}}=1.1,C_{{t4}}=2
\end{align*}
where turbulent eddy viscosity 
is defined as $\nu_{t}$ and the stress tensor is $S$,with $\Omega$ being the rotational tensor,
$d$ is the distance from the closest surface and $\Delta U^{2}$ is the norm of the difference between the velocity. 
The model defines a transport equation for a new viscosity-like variable $\tilde {\nu}$. To solve the SA equation for the turbulence closure of RANS, the following boundary conditions are prescribed; at wall $\tilde{\nu} = 0$, freestream/inlet $\tilde{\nu} = 3 \nu$ and  outlets are defined with Neumann condition. In the present work we use the SA closure to generate the steady state field of $\nu_{t}$ by solving the RANS equation, which is used as training data for the developing machine learning model. In addition, we also utilize the two-equation RNG $k-\epsilon$ \cite{yakhot1992development} and $k-\omega$ SST \cite{menter1994two} models for demonstrating the generality of the proposed framework.   These model equations are widely used for a variety of practical turbulence modeling applications for simulating complex flows. We also generate the steady state $\nu_{t}$ field using these models. 
The equations for these models are omitted here for brevity but correspond to Eqs. (4.35), (4.42) and (5.10) for RNG $k-\epsilon$ in \cite{yakhot1992development} and Eqs. A1-2, A13-15 for $k-\omega$ SST in \cite{menter1994two}. Using OpenFOAM-based nomenclature, Table \ref{TableA} provides a summary of the boundary conditions for each closure model employed during the data-generation phase of our workflow.\\
Once our governing equations are formulated, they must be solved on a discrete grid. We utilize the finite-volume method to ensure local conservation of our governing laws. This method requires the numerical calculation of spatial gradients for which we use a second-order accurate discretization. In particular, our viscous terms use a purely central discretization, whereas the advective terms use the LUST (blended linear and linear upwind scheme) to control for any spurious Gibbs phenomena. We note that these discretization methods are commonly used in practical CFD applications. Our problem utilizes a steady-state solver given by the SIMPLE algorithm. 

We built our finite volume solver and machine learning deployment framework by integrating the C-backend of Tensorflow 1.14 \cite{tensorflow2015-whitepaper} into OpenFOAM 5.0. 



\vspace{0.5cm}

\begin{table}
\centering
\footnotesize
\resizebox{\textwidth}{!}{\begin{tabular}{llllllllll}
\multicolumn{10}{c}{OpenFOAM Boundary Conditions}                                                                             \\ 
\cline{1-10}
\\
\multicolumn{10}{c}{SA}                                                                                \\
\cline{1-10}
\multicolumn{4}{l|}{Boundary Type} & \multicolumn{3}{c}{$\nu_{t}$} & \multicolumn{3}{c}{$\tilde{\nu}$}             \\
\cline{1-10}
\multicolumn{4}{l|}{Inlet}         & \multicolumn{3}{c}{\texttt{calculated}}            & \multicolumn{3}{c}{$3\nu$}             \\ 
\multicolumn{4}{l|}{Outlet}        & \multicolumn{3}{c}{\texttt{calculated}}            & \multicolumn{3}{c}{\texttt{zeroGradient}}             \\ 
\multicolumn{4}{l|}{Wall}          & \multicolumn{3}{c}{\texttt{fixedValue}$=0.0$}            & \multicolumn{3}{c}{0.0}             \\
\cline{1-10}
\\
\multicolumn{10}{c}{$k-\omega$ SST}                                                                     \\ 
\cline{1-10}
\multicolumn{4}{l|}{Boundary Type} & \multicolumn{2}{c}{$k$} & \multicolumn{2}{c}{$\omega$} & \multicolumn{2}{c}{$\nu_t$} \\
\cline{1-10}
\multicolumn{4}{l|}{Inlet}         & \multicolumn{2}{c}{\texttt{fixedValue}} & \multicolumn{2}{c}{\texttt{fixedValue}} & \multicolumn{2}{c}{\texttt{calculated}} \\
\multicolumn{4}{l|}{Outlet}        & \multicolumn{2}{c}{\texttt{zeroGradient}} & \multicolumn{2}{c}{\texttt{zeroGradient}} & \multicolumn{2}{c}{\texttt{calculated}} \\
\multicolumn{4}{l|}{Wall}          & \multicolumn{2}{c}{\texttt{kqRwallFunction}} & \multicolumn{2}{c}{\texttt{omegaWallFunction}} & \multicolumn{2}{c}{\texttt{nutUSpaldingWallFunction}} \\
\cline{1-10}
\\
\multicolumn{10}{c}{$k-\epsilon$ RNG}                                                                     \\ 
\cline{1-10}
\multicolumn{4}{l|}{Boundary Type} & \multicolumn{2}{c}{$k$} & \multicolumn{2}{c}{$\epsilon$} & \multicolumn{2}{c}{$\nu_t$} \\
\cline{1-10}
\multicolumn{4}{l|}{Inlet}         & \multicolumn{2}{c}{\texttt{fixedValue}} & \multicolumn{2}{c}{\texttt{fixedValue}} & \multicolumn{2}{c}{\texttt{calculated}} \\
\multicolumn{4}{l|}{Outlet}        & \multicolumn{2}{c}{\texttt{zeroGradient}} & \multicolumn{2}{c}{\texttt{zeroGradient}} & \multicolumn{2}{c}{\texttt{calculated}} \\
\multicolumn{4}{l|}{Wall}          & \multicolumn{2}{c}{\texttt{kqRwallFunction}} & \multicolumn{2}{c}{\texttt{epsilonWallFunction}} & \multicolumn{2}{c}{\texttt{fixedValue=0.0}} \\
\cline{1-10}
\end{tabular}}\\
\caption{OpenFOAM-based nomenclature to describe the boundary conditions for each turbulence model employed in this study.} 
\label{TableA}
\end{table}

\subsection{ML-RANS Integration}\label{subsection:ml-rans}

The core idea behind this investigation motivates the development of a surrogate turbulent-eddy viscosity model to bypass the solution of any extra equations for closure. Through this we aim to obtain computational speed-up by reducing the number of PDEs to solve iteratively to steady-state. While the previous section has introduced the SA equation as our reference model, we note that models of any fidelity can be used to generate the data needed by the framework. This shall also be demonstrated on two examples of two-equation models below. The core procedure is given below:

\begin{enumerate}
    \item \underline{Data generation phase}
    \begin{itemize}
        \item Select numerical experiment locations in the design space. In our study, this corresponds to different boundary conditions for inlet velocity and different step heights.
        \item Initialize initial conditions for numerical experiments using low fidelity approximations (such as potential flow).
        \item Generate steady-state turbulent eddy-viscosity profiles using an \emph{a priori} specified closure model (in this case SA).
        \item Augment initial condition information through feature preprocessing or geometric information embedding (for instance distance from the wall). 
        \item Save training data in the form of pointwise input-output pairs of initial conditions and steady-state turbulent eddy viscosities.
    \end{itemize}
    
    \vspace{0.2cm}

    \item \underline{Training phase}
    \begin{itemize}
        \item Train an ML surrogate to predict the pointwise steady-state turbulent eddy viscosity field given low fidelity initial conditions as input. Optionally, the ML surrogate may also be given explicit information about the control parameter.
        \item Train framework using 90\% of the total data set, referred to as the training data; keeping the rest (10\%) as validation data for model selection and to detect any potential overfitting of the model. {Training and validation data are both comprised of samples from all control parameters.}
        \item Save the trained model for deployment.
    \end{itemize} 
    
    \vspace{0.2cm}
    
    \item \underline{Testing phase}
    \begin{itemize}
        \item Choose a new design point in experiment space, for instance, a new value for inlet velocity or step height of the backward facing step {which was not present in the training data}.
        \item Generate initial conditions for this new point using low fidelity initial condition - these will be the \emph{test} inputs to the trained framework. Augment inputs with control parameter information if needed.
        \item Deploy the previously trained framework to predict an approximation for the steady-state turbulent eddy viscosity and fix this quantity permanently.
        \item Solve the Reynolds-averaged pressure and velocity equations to steady-state while utilizing the fixed steady-state turbulent eddy viscosity.
        \item Assess the impact of the surrogate eddy viscosity through quantitative metrics.
    \end{itemize}
\end{enumerate}

We now proceed with a tabulation of our input and output features in this study. Our inputs identify the region in space for model deployment through inputs of the finite-volume cell-centered coordinates and the initial conditions. We note that our inputs and outputs were scaled to having unit mean and zero variance for each feature to enable easier training. In this particular study we undertake two experiments. In our first study, we train an ML surrogate given by
\begin{align}
    \mathbb{M}_1 : u^p (\textbf{x}), v^p (\textbf{x}), x^c (\textbf{x}), y^c (\textbf{x}) \rightarrow \nu_t (\textbf{x})
\end{align}
and interpolate between inlet velocities. Here $u^p$ and $v^p$ indicate the horizontal and vertical velocities resulting from the initial conditions and where $x^c$, $y^c$ are the cell-centered coordinates of the grid in the domain. The output, $\nu_t(\textbf{x})$, is the steady-state turbulent eddy viscosity potentially obtained from any choice of RANS closure strategies.  Our second study interpolates between different geometries (by varying the step height). For this purpose we establish a map that is augmented with control parameter information as follows
\begin{align}
    \mathbb{M}_2 : u^p (\textbf{x}), v^p (\textbf{x}), x^c (\textbf{x}), y^c (\textbf{x}), h \rightarrow \nu_t (\textbf{x})
\end{align}
where our original set of input features have been augmented by the step height $h$. 

\begin{remark}
{The primary motivation for using potential flow solution quantities as input features stems from a desire to have low-fidelity features (that are easily computable) for training the ML map. This aligns with several ML-based emulation studies where low-fidelity models are used to reconstruct the influence of unavailable fidelity (for example in RANS \cite{ling2016reynolds,wang2017physics,wu2018physics,matai2019zonal,sotgiu2019towards,taghizadeh2020turbulence} and large eddy simulation \cite{gamahara2017searching,maulik2017neural,maulik2019sub} subgrid modeling). The use of potential flow is one such approach for constructing a map but it is not the sole approach we espouse. Any underlying information about the system may be utilized for training the $\nu_t$ emulator provided the learning is accurate and generalizable.}
\end{remark}

Both our surrogate maps $\mathbb{M}_1$ and $\mathbb{M}_2$ are given by a neural network with 6 hidden layers and 40 neurons in each layer. We use a {tangent sigmoidal} activation for the hidden layers while the output layer uses a linear activation. In our training procedure we use the ADAM optimizer\cite{kingma2014adam} with a learning rate of 0.001. For $\mathbb{M}_1$, the fully trained network achieves a coefficient of determination\footnote{$R^2 \equiv 1 - \frac{\sum_i  (y_i - y_i^{\rm{pred}})^2}{\sum_i(y_i - \bar{y})^2} $ for data $y_i$, predictions $y_i^{\rm{pred}}$ and mean $\bar{y}$. } of $R^2=0.998$ for both training and validation data sets indicating a successful parameterization. For $\mathbb{M}_2$, we obtain a similarly high accuracy of $R^2 = 0.997$. The training for both networks was terminated using an early-stopping criterion. If validation errors did not improve for 10 epochs, the training would exit with the best model (corresponding to the lowest validation loss until then). We note that, in deployment, we truncated negative value predictions, akin to the built-in limiters of the SA model itself. We note that the hyperparameters of our network and its architecture were hand-tuned for accuracy due to its relative simplicity. Further analysis of this training will be performed in Section \ref{ML_Section}.

\begin{remark}
{The utilization of a tangent sigmoidal (tanh) activation ensures that $\nu_t$ predictions are differentiable with respect to the input mesh coordinates. While deep learning methods have demonstrated success with rectified linear (ReLU) activation to solve the vanishing gradient problem \cite{hochreiter1998vanishing}, the smooth transformation obtained via tanh aids us in avoiding discontinuous $\nu_t$ profiles from piece-wise linear reconstructions obtained from ReLU.}
\end{remark}

\section{Results}

In the following section, we outline our machine learning effectiveness through \emph{a priori} analyses followed by a discussion of the results from its deployment as a surrogate model.

\subsection{Machine learning}
\label{ML_Section}

\begin{figure}
    \centering
    \mbox{
    \subfigure[$\mathbb{M}_1$ Convergence]{\includegraphics[width=\textwidth]{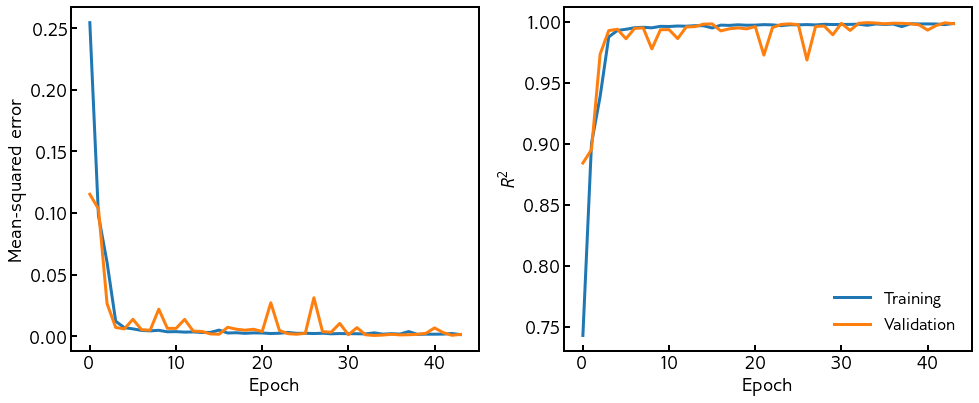}}
    } \\
    \mbox{
    \subfigure[$\mathbb{M}_2$ Convergence]{\includegraphics[width=\textwidth]{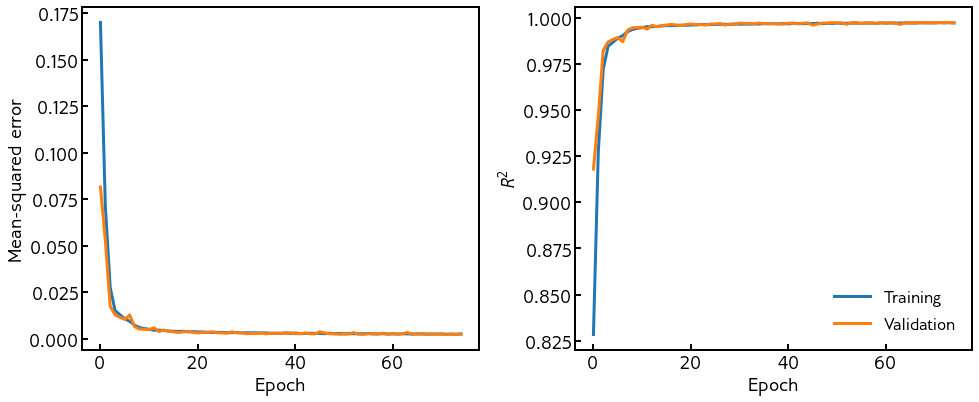}}
    }
    \caption{Convergence while learning the machine learning surrogate $\mathbb{M}_1$ (top) and $\mathbb{M}_2$ (bottom).}
    \label{Fig_ML_Training}
\end{figure}

We proceed by outlining the effectiveness of the machine learning through statistical estimates for its performance. Figure \ref{Fig_ML_Training} shows the progress to convergence for our learning frameworks. It was observed that approximately 40 epochs were sufficient for an accurate parameterization of the input-output relationship after which the early-stopping criteria terminated training. The training data set for $\mathbb{M}_1$ consisted of 205,390 samples whereas $\mathbb{M}_2$ consisted of 808,882 samples. The reader may note here, that while $\mathbb{M}_2$ was sampled across only 6 different step heights (compared to 10 different boundary velocities for $\mathbb{M}_1$), the varying geometries for each step height contributed to a greater number of training points in total. Out of the total number of samples 90\% were retained for the purpose of training while the rest were utilized for validation. 

\begin{figure}
    \centering
    \mbox
    {
    \subfigure[$\mathbb{M}_1$ Scatter ]{\includegraphics[width=0.5\textwidth]{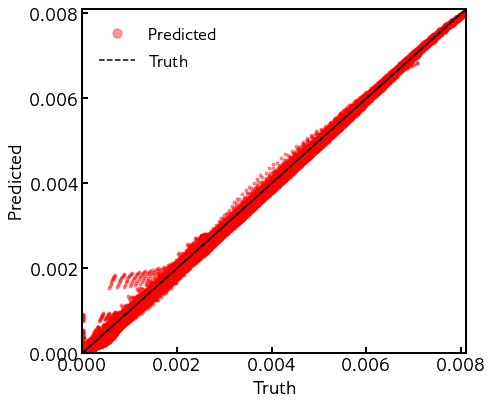}}
    \subfigure[$\mathbb{M}_1$ Density]{\includegraphics[width=0.5\textwidth]{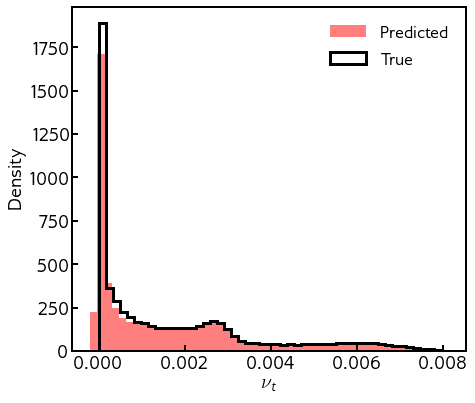}}
    } \\
    \mbox
    {
    \subfigure[$\mathbb{M}_2$ Scatter ]{\includegraphics[width=0.5\textwidth]{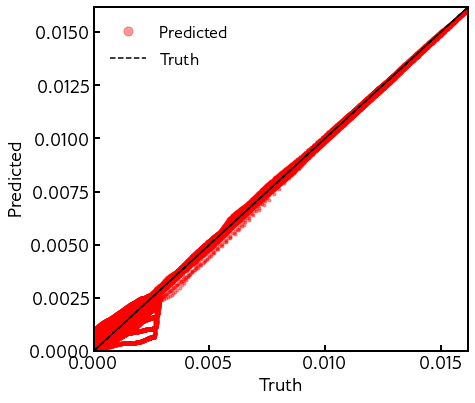}}
    \subfigure[$\mathbb{M}_2$ Density]{\includegraphics[width=0.5\textwidth]{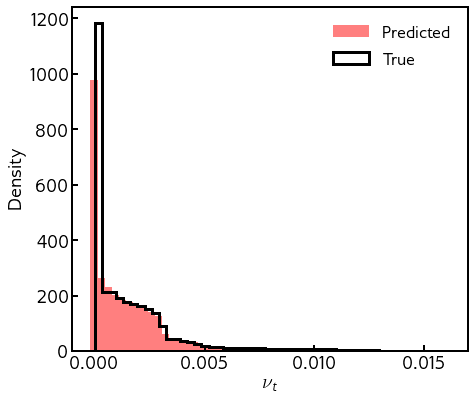}}
    } \\
    \caption{An \emph{a priori} statistical assessment of the trained surrogate models $\mathbb{M}_1$ (top) and $\mathbb{M}_2$ (bottom) for learning $\nu_t$.}
    \label{Fig_ML_Assessment}
\end{figure}

A statistical assessment of the learning is shown in Figure \ref{Fig_ML_Assessment}. We note that these plots utilize training and validation data for all simulations for the purpose of \emph{a priori} assessment. The scatter plots show a good agreement between modeled and true magnitudes of $\nu_t$ particularly for the higher magnitudes. Some deviation is observed at the lower magnitudes potentially due to slight inaccuracies in the characterization of the transition from the separation zone to the free-stream. Notice how the probability density for $\mathbb{M}_1$ has a much smaller maximum value than $\mathbb{M}_2$ thus indicating a narrower spread of data. This may be explained by a wider spread of sampled $\nu_t$ in our training data set for the latter. 

The probability distribution plots for the predicted and true models also show good agreement. We note that the $\nu_t$ magnitude are captured accurately by the predictions. As a further exploration of the benefits of using a deep learning framework, the \emph{a priori} performance of the surrogate $\mathbb{M}_1$ is assessed against linear and polynomial regression using the same inputs and output. To account for on-node memory limitations for this study, we truncated the polynomial order to 6. The $R^2$ values (for validation data) for these assessments are shown in Table \ref{Table2}. The proposed framework can be seen to outperform the linear and polynomial regression methods successfully. {At this point, it is important to note that the number of trainable parameters of polynomial methods are quite lower than the DNN, particularly at lower orders of approximation. While this might imply computational competitiveness for this particular test case, polynomial models often suffer from the bias-variance trade-off where increasing the order leads to very poor generalization. This can be observed with gradual increase in the polynomial order where the validation $R^2$ is seen to saturate (to a value less than that obtained by the DNN) and drop suddenly if overfitting has started. In addition, while DNNs may be over-parameterized, they may readily be implemented by using freely available optimized machine learning libraries which take advantage of specialized computer hardware such as graphical and tensor processing units for acceleration.}



\begin{table}
\scriptsize
\centering
\begin{tabular}[b]{cccccccccccccc}
      \multicolumn{12}{c}{Validation-$R^2$} \\
      \hline
      Order & 1 & 2 & 3 & 4 & 5 & 6 & 7 & 8 & 9 & 10 & DNN\\ \hline
      $\mathbb{M}_1$ & 0.146 & 0.249 & 0.421 & 0.745 & 0.861 & 0.937 & 0.959 & 0.972 & 0.981 & 0.967 & 0.998 \\ \hline
      $\mathbb{M}_2$ & 0.266 & 0.419 & 0.548 & 0.805 & 0.915 & 0.949 & 0.969 & 0.959 & 0.915 &  0.031 & 0.998 \\ \hline
      \vspace{0.2cm}
\end{tabular}

\begin{tabular}[b]{cccccccccccccc}
      \multicolumn{12}{c}{Number of parameters} \\
      \hline
      Order & 1 & 2 & 3 & 4 & 5 & 6 & 7 & 8 & 9 & 10 & DNN\\ \hline
      $\mathbb{M}_1$  & 5 & 15 & 35 & 70 & 126 & 210 & 330 & 495 & 715 & 1001 & 10,081\\ \hline
      $\mathbb{M}_2$  & 6 & 21 & 56 & 126 & 252 & 462 & 792 & 1287 & 2002 & 3003 & 10,121\\ \hline
\end{tabular}
\caption{Validation $R^2$ values (top) and number of trainable parameters (bottom) for different order polynomial fits for the proposed experiments. The proposed deep learning framework (DNN) is seen to outperform the polynomial regressions. It was also observed that RANS deployments did not converge to the preset solver tolerances when using the linear and polynomial regressions.}
\label{Table2}
\end{table}

\subsection{Generalization across boundary conditions}

In the following sections, we outline the results from our proposed formulation for an interpolation task within a range of control parameters sampled for training data by the SA turbulence model. We remind the reader that the machine learning framework predicts the steady-state viscosity when the simulation is initialized and the solver then utilizes this fixed viscosity for its progress to convergence. 

We first outline results from the evaluation of surrogate $\mathbb{M}_1$. We shall be deploying our proposed framework on two \emph{testing} situations with inlet velocity conditions given by 44.2 m/s (denoted S1) and 49.5 m/s (denoted S2). Both these inlet velocity conditions constitute data that is unseen by the network during training. Our training data set consists of steady-state $\nu_t$ magnitudes obtained from using Spalart-Allmaras on inlet velocities of 40, 41, 42, 43, 44, 45, 46, 47, 48 and 49 m/s.

\begin{figure}
    \centering
    \includegraphics[width=\textwidth]{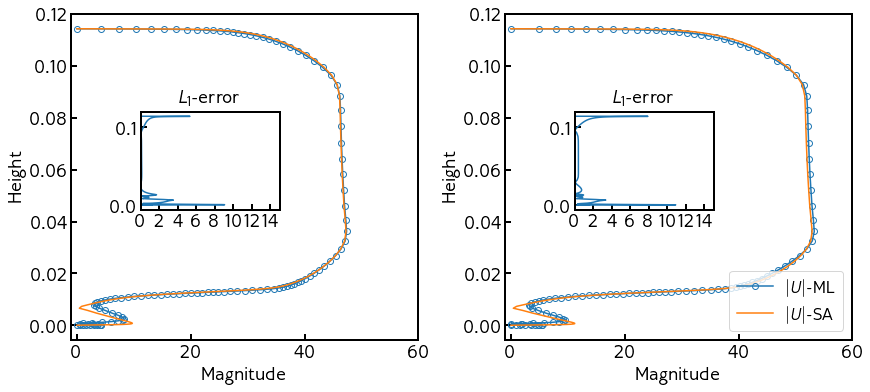}
    \caption{$|U|$ predictions (in m/s) at probe location 1 for surrogate deployment S1 (left) and S2 (right) using data-driven map $\mathbb{M}_1$.}
    \label{Fig_Lineplot_1}
\end{figure}

\begin{figure}
    \centering
    \includegraphics[width=\textwidth]{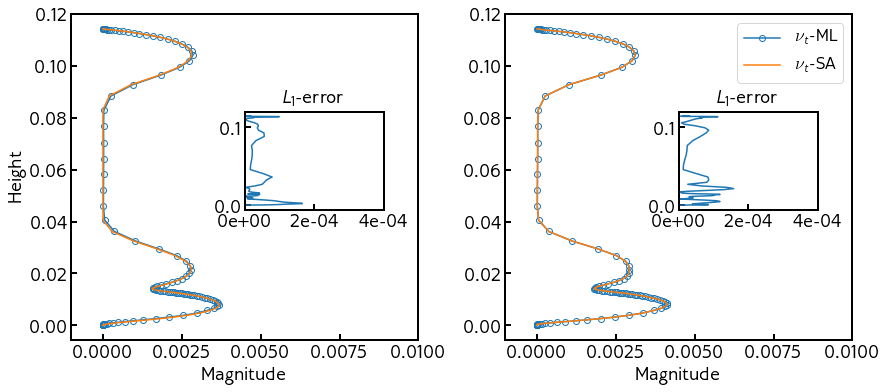}
    \caption{Turbulent eddy-viscosity predictions (in m\textsuperscript{2}/s) at probe location 1 for surrogate deployment S1 (left) and S2 (right) using data-driven map $\mathbb{M}_1$.}
    \label{Fig_Lineplot_2}
\end{figure}

Figure \ref{Fig_Lineplot_1} shows a line plot for the velocity magnitudes ($|U|$) (for both S1 and S2) at probe location 1 (previously defined in Section \ref{Problem}). A close agreement is observed between the converged simulation using the machine learning surrogate and the standard deployment of SA. A strength of this proposed approach is that the velocity and pressure solvers were untouched during our surrogate modeling and the converged fields preserve their respective symmetries. Figure \ref{Fig_Lineplot_2} shows the performance of the network in predicting the output quantity (i.e., $\nu_t$) at probe location 1. {The data-driven map is seen to capture the $\nu_t$ profiles accurately.}

\begin{figure}
    \centering
    \includegraphics[width=\textwidth]{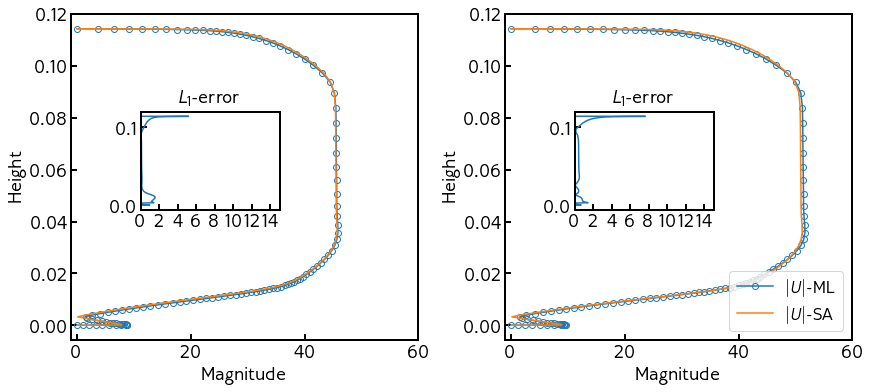}
    \caption{$|U|$ predictions (in m/s) at probe location 2 for surrogate deployment S1 (left) and S2 (right) using data-driven map $\mathbb{M}_1$.}
    \label{Fig_Lineplot_3}
\end{figure}

\begin{figure}
    \centering
    \includegraphics[width=\textwidth]{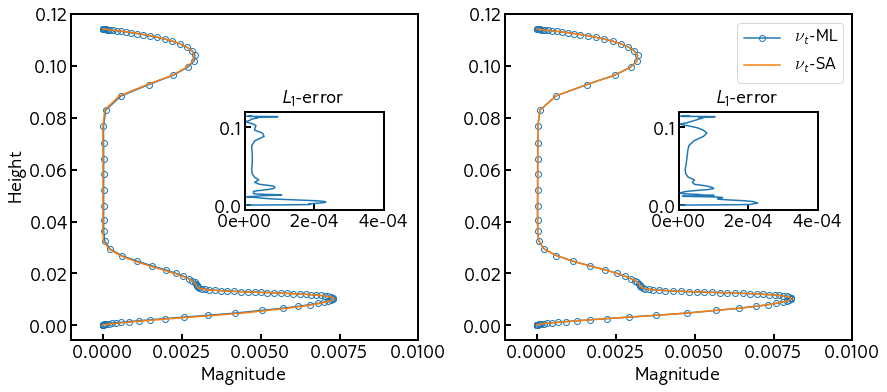}
    \caption{Turbulent eddy-viscosity predictions (in m\textsuperscript{2}/s) at probe location 2 for surrogate deployment S1 (left) and S2 (right) using data-driven map $\mathbb{M}_1$.}
    \label{Fig_Lineplot_4}
\end{figure}

\begin{figure}
    \centering
    \includegraphics[width=\textwidth]{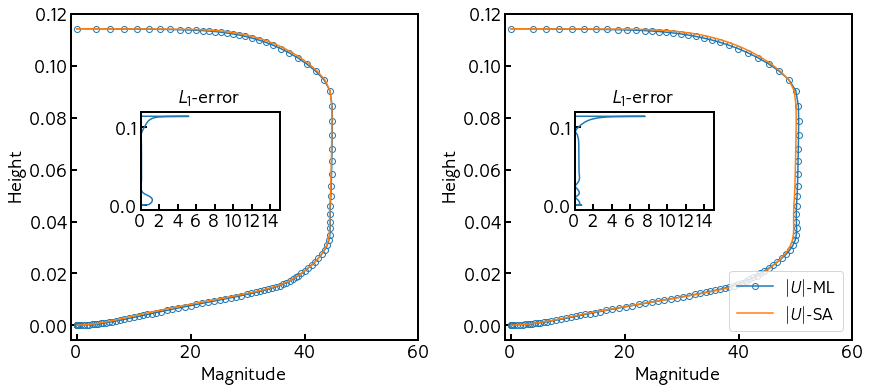}
    \caption{$|U|$ predictions (in m/s) at probe location 3 for surrogate deployment S1 (left) and S2 (right) using data-driven map $\mathbb{M}_1$.}
    \label{Fig_Lineplot_5}
\end{figure}

\begin{figure}
    \centering
    \includegraphics[width=\textwidth]{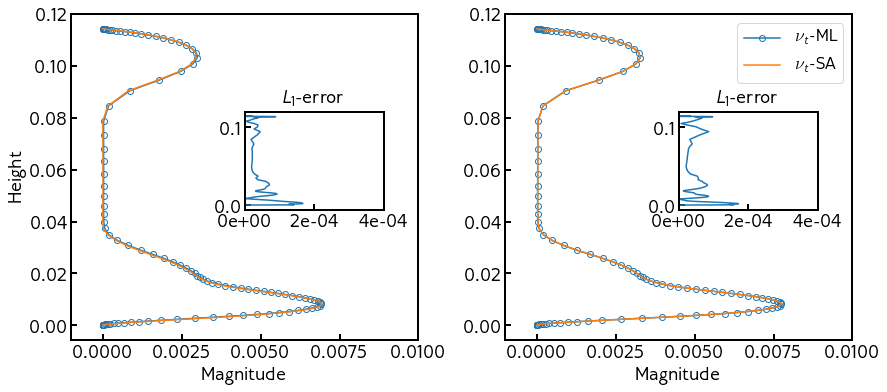}
    \caption{Turbulent eddy-viscosity predictions (in m\textsuperscript{2}/s) at probe location 3 for surrogate deployment S1 (left) and S2 (right) using data-driven map $\mathbb{M}_1$.}
    \label{Fig_Lineplot_6}
\end{figure}

\begin{figure}
   \centering
    \mbox{
    \subfigure[$\nu_t$ $L_1$-error]{\includegraphics[trim={0 6cm 0 0},clip,width=0.48\textwidth]{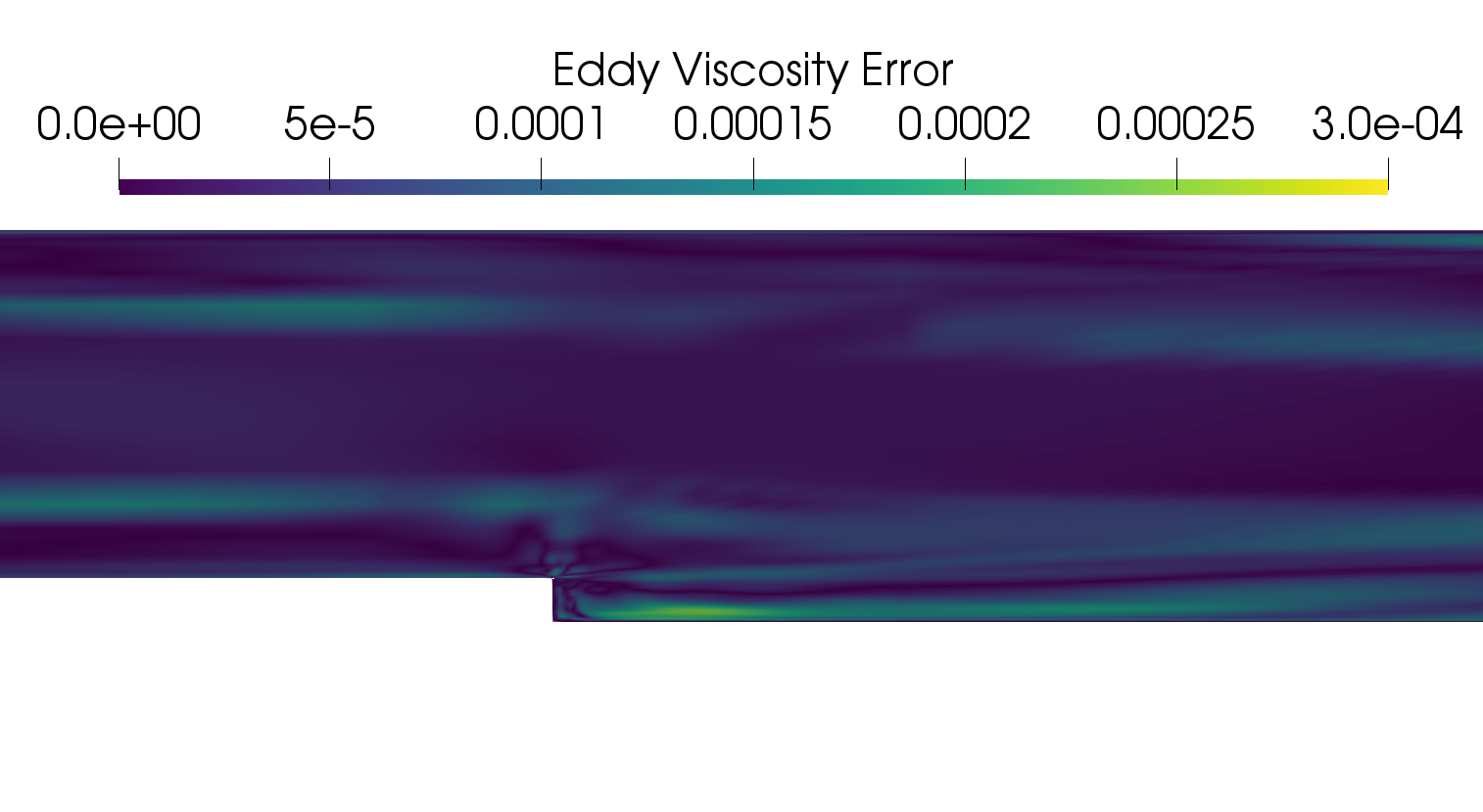}}
    \subfigure[$|U|$ $L_1$-error]{\includegraphics[trim={0 6cm 0 0},clip,width=0.48\textwidth]{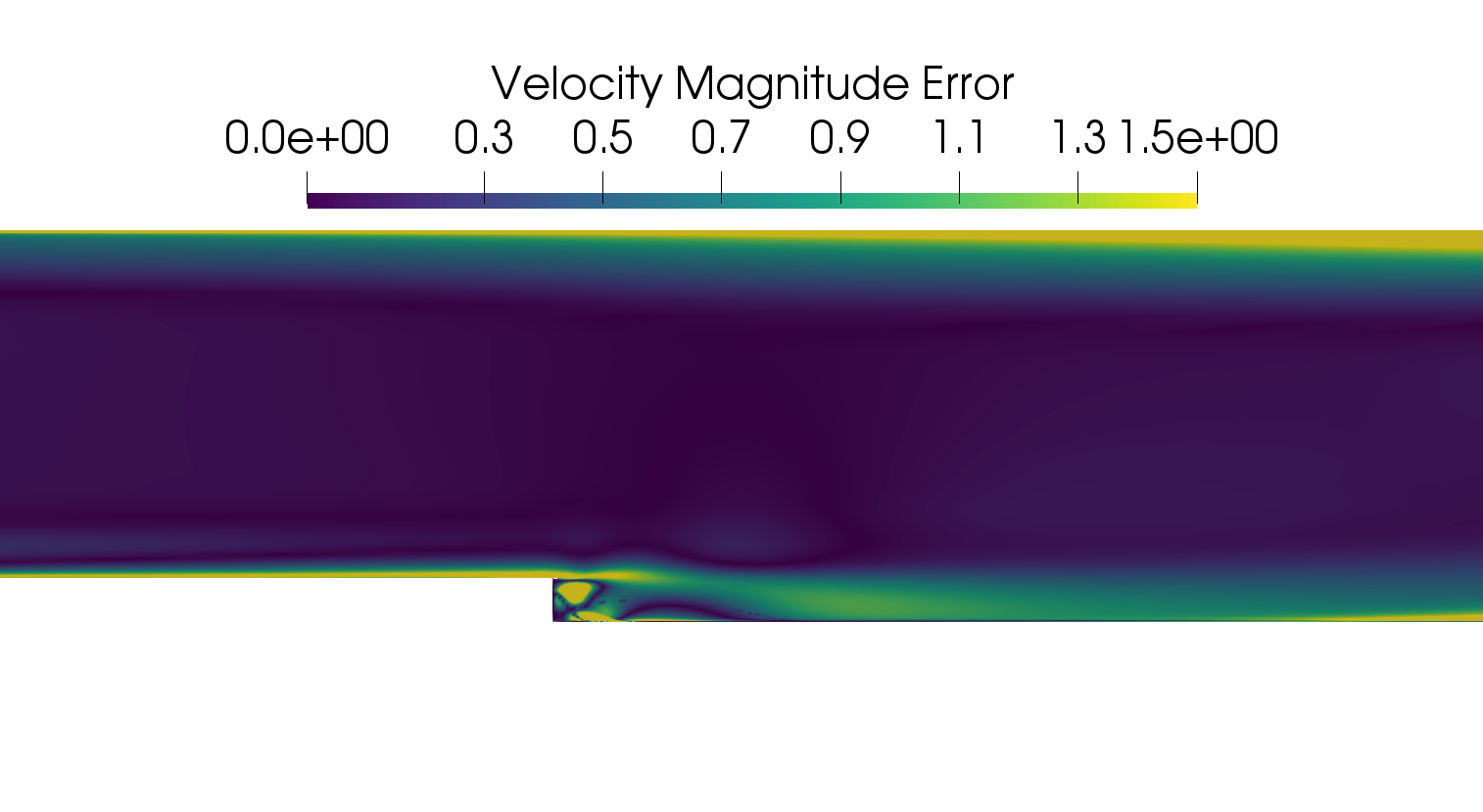}}
    }
    \caption{$L_1$ errors for deploying the ML framework on test-case S1 using data-driven map $\mathbb{M}_1$. Errors are concentrated near the region of the step, due to mesh anisotropy as well as complicated physics. In addition, error spikes are seen on the upper boundary. Note that error magnitudes are an order of magnitude less than the quantities of interest.}
    \label{Fig_Contour_2}
\end{figure}

Further assessments of the surrogate are performed for both our testing simulations at different probe locations for $|U|$ and $\nu_t$. As shown in Figure \ref{Fig_Lineplot_3} and Figure \ref{Fig_Lineplot_4}, a good agreement between the one-equation model and the surrogate is once again observed downstream of the step. Our third probe location (located before the flow reaches the step) also shows that the deep learning framework approximates the behavior of SA appropriately, as shown in Figures \ref{Fig_Lineplot_5} and \ref{Fig_Lineplot_6}. Finally, Figure \ref{Fig_Contour_2} shows $L_1$ errors between truth and prediction where it is observed that errors are an order of magnitude lower than the quantities of interest. 

\begin{figure}
    \centering
    \mbox{
    \includegraphics[width=0.48\textwidth]{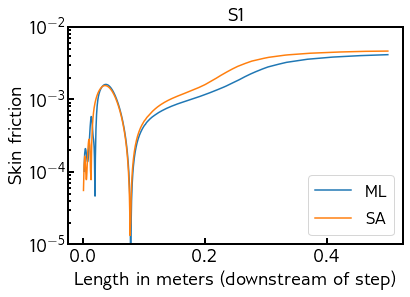}
    \includegraphics[width=0.48\textwidth]{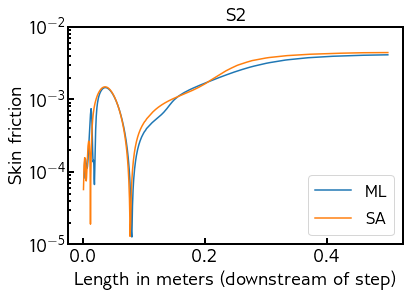}
    }
    \caption{Skin-friction coefficient predictions downstream of the step for deploying the ML framework on test-cases S1 (left) and S2 (right) using data-driven map $\mathbb{M}_1$. The ML model is seen to over-estimate flow reattachment length by a small amount.}
    \label{Fig_CF_1}
\end{figure}

Finally, we assess the accuracy of the proposed framework in terms of skin friction prediction in Figure \ref{Fig_CF_1}. The skin-friction is defined as 
\begin{align}
    C_f = \frac{\tau_w}{\frac{1}{2} \rho U^2},
\end{align}
where $\tau_w = \nu_{eff} \frac{\partial \bar{u}_1}{\partial y}$ is the wall shear stress, $\nu_{eff} = \nu_t + \nu$ is the effective kinematic viscosity and $U$ is the free-stream inlet velocity (as introduced previously). The skin-friction coefficient may be used to determine the reattachment point of the flow after it has separated over the step. This is manifested by a tendency for $C_f$ to become zero at a certain distance downstream of the step indicating that the flow has touched the lower wall again. {Our proposed framework with $\mathbb{M}_1$ is seen to predict the reattachment location well. However, slight deviations are observed downstream of the reattachment before the skin-friction profile matches that of the PDE-based model.} This may be explained by point-wise predictions for $\nu_t$ which may lead to residual noise in the directly predicted `steady-state' turbulent eddy-viscosity.

\subsubsection{Speedup from the proposed approach}

\begin{figure}
    \centering
    \includegraphics[width=\textwidth]{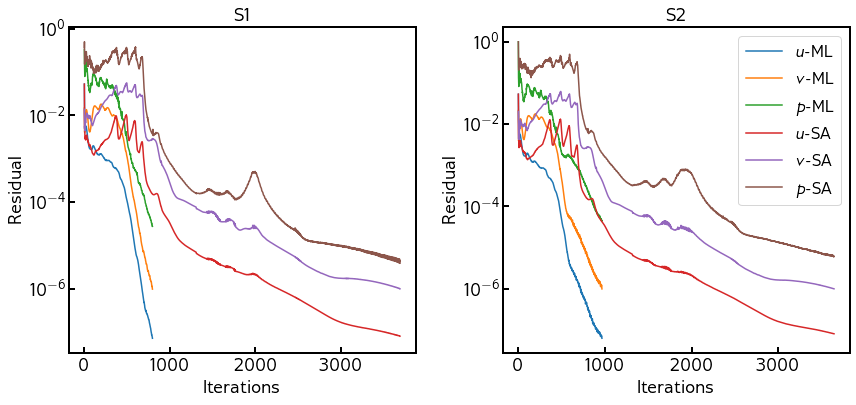}
    \caption{Residual plots for deploying the ML framework on test-cases S1 (left) and S2 (right) exhibiting the speedup from the proposed approach using data-driven map $\mathbb{M}_1$.}
    \label{Fig_Lineplot_7}
\end{figure}

One of the advantages of the proposed formulation is the removal of an additional PDE for the calculation of $\tilde{\nu}$ utilized for specifying $\nu_t$. While one might expect a speedup due to the reduced dimensionality of the coupled PDE system, it is not clear if this would manifest in a reduced time to solution due to the complicated interplay with deep learning prediction errors and the steady-state solvers for velocity and pressure. Through an empirical assessment, we determined that the proposed methodology would allow for considerably \emph{larger} relaxation factors for the steady-state velocity and pressure solvers. In the case of the two surrogate models tested above, relaxation factors of 0.9 for both pressure and velocity led to a converged solution in 687 iterations of the steady-state solver for S1 and 716 iterations for S2. The SA implementations, however, needed 3696 and 3646 iterations for convergence while utilizing relaxation factors of 0.5 for pressure, 0.9 for velocity and 0.3 for $\tilde{\nu}$. We note that the relaxation factors for this deployment were hand-tuned to obtain convergence and that speedup factors are relative. A graphical representation of the speedup is provided in Figure \ref{Fig_Lineplot_7}. In terms of time to solution for experiment S1, the proposed framework required 14.49 seconds for convergence whereas the SA model required 102.76 seconds. For experiment S2, the corresponding times-to-solution were 15.08 and 112.60 seconds. We note that all experiments were performed using a serial execution of OpenFOAM on an Intel Core-i7 processor with a clockspeed of 1.90 GHz.

\begin{remark}
{It must be noted that the current framework accelerates the solution of the system to a different end-result, as observed in Figure \ref{Fig_CF_1}. An accurate data-driven map for the turbulent eddy viscosity may lead to a small deviation of quantities of interest between the true and reduced-order solution. This is different from methods that accelerate the solution of the system to the same end-result (for example by using preconditioners, multigrid methods or Krylov solvers).}
\end{remark}

\subsection{Generalization across geometries}

In the following section, we assess the utility of the proposed surrogate model, $\mathbb{M}_2$, for predicting the steady-state turbulent eddy viscosity across mesh configurations \textcolor{black}{instead of across boundary conditions}. A series of tests are performed to assess the viability of the methodology for generalization across geometries. These are performed by generating training data for different step heights \textcolor{black}{of 0.5$h$, 0.75$h$, 1.25$h$, 1.5$h$, 1.75$h$, 2.0$h$ (and therefore different meshes)} and assessing interpolation between them. \textcolor{black}{We fix the inlet velocity at 44.2 m/s to assess generalization across geometries within computational constraints. A parameter sweep across both varying geometries and boundary conditions would require more full-order solves - but the conclusions from this study would be valid regardless}. Recall that to set up a surrogate modeling framework for $\nu_t$ in this context, our machine learning framework is adapted to allow for the step height as an explicit feature input in addition to the initial conditions and mesh coordinate points. {We then test the trained model for predicting the eddy viscosity profile for a value of step height ($1.9h$) held out from the training data set.} 

\begin{figure}
    \centering
    \mbox{
    \subfigure[Location 1]{\includegraphics[width=0.33\textwidth]{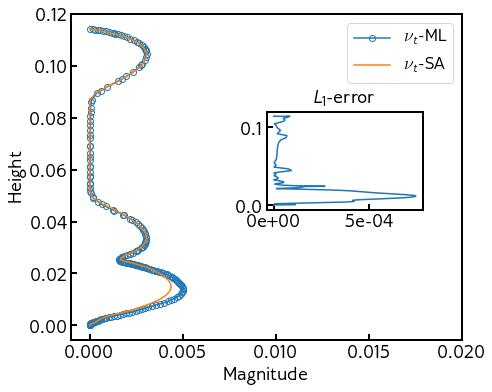}}
    \subfigure[Location 2]{\includegraphics[width=0.33\textwidth]{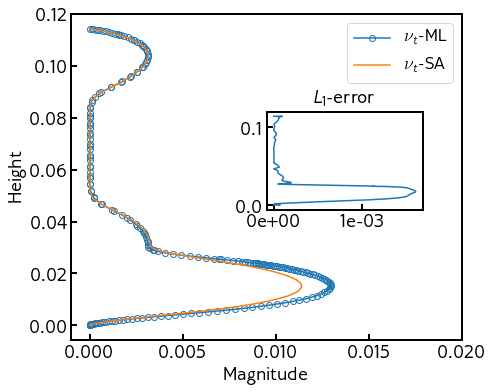}}
    \subfigure[Location 3]{\includegraphics[width=0.33\textwidth]{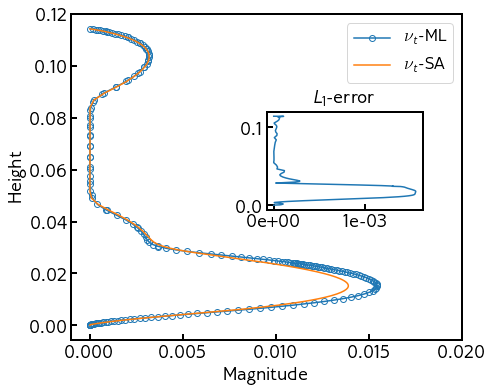}}
    }
    
    \mbox{
    \subfigure[Location 1]{\includegraphics[width=0.33\textwidth]{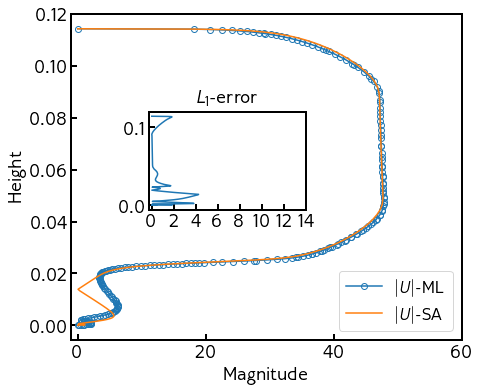}}
    \subfigure[Location 2]{\includegraphics[width=0.33\textwidth]{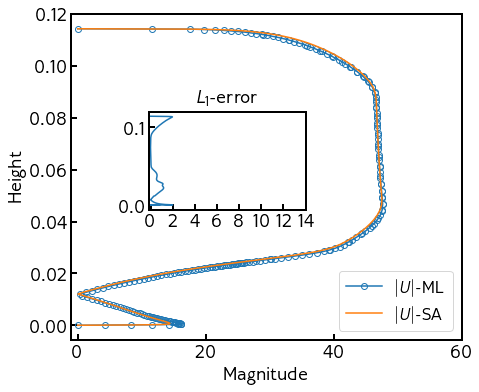}}
    \subfigure[Location 3]{\includegraphics[width=0.33\textwidth]{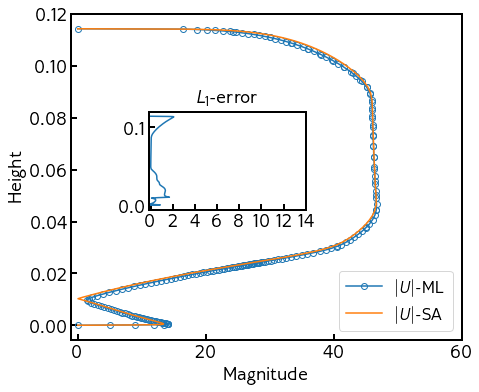}}
    }

    \caption{Predictive capability of ML framework for unseen geometries (i.e., data-driven map $\mathbb{M}_2$). The assessments in these figures are for a case where the backward facing step with height $1.9h$ was \emph{not} a part of the training data set. Instead, training data was generated from step heights of 0.5$h$, 0.75$h$, 1.25$h$, 1.5$h$, 1.75$h$, 2.0$h$. The figure shows line plots of $\nu_t$ (top) and velocity magnitude (bottom) for three different probe locations.}
    \label{fig:dg_pred_2}
\end{figure}

\begin{figure}
    \centering
    \mbox{
    \subfigure[Standard Mesh]{\includegraphics[width=0.48\textwidth]{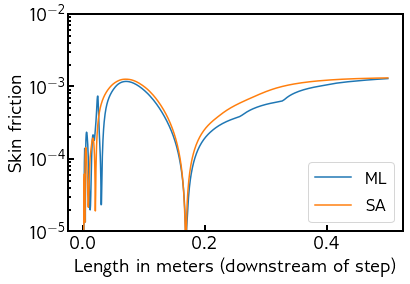}}
    \subfigure[Fine Mesh]{\includegraphics[width=0.48\textwidth]{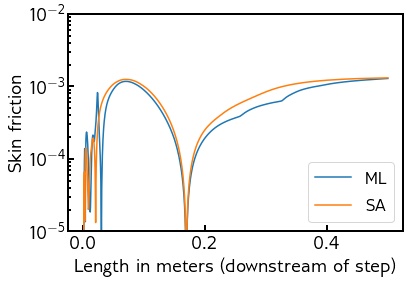}}
    }
    \caption{Skin-friction coefficient predictions downstream of the step for deploying the ML framework on a test step height of $1.9h$ using data-driven map $\mathbb{M}_2$. {The finer mesh possessed approximately twice the degrees of freedom as the standard mesh.}}
    \label{Fig_CF_2}
\end{figure}

The results for the $\nu_t$ and $|U|$ profiles at three locations are shown in Figure \ref{fig:dg_pred_2}. For $\nu_t$ magnitudes, the ML framework is able to predict accurately near the top wall but is prone to overestimating the magnitude on the lower wall. However, we draw attention to the fact that trained network is able to predict a larger magnitude of $\nu_t$ (associated with true predictions from SA for a larger step height) successfully. {Also, the framework successfully reproduces the qualitative trends in the $\nu_t$ profiles on the lower wall with increasing distance from the step.} For velocity magnitudes, the resulting steady-state solution using the ML-model leads to a mismatch in the behavior for location 1 (within the recirculation region). However, accurate results in location 2 and location 3 are observed. For instance, successfully learning $\mathbb{M}_2$ allows for the framework to accurately predict that location 3 will still have recirculation for this step height, unlike with height $h$. {Subsequently, we assess skin-friction predictions for the test step-height in Figure \ref{Fig_CF_2} (a). The performance is seen to be similar to that observed for the previous set of assessments. The reattachment location is captured well, and slight deviations are observed downstream of the reattachment before the skin-friction profile matches that of the PDE-based model. We also perform a test for assessing the deployment of trained models at varying mesh densities. This test utilizes a geometry of height $1.9h$ but utilizes a mesh with 292,214 degrees of freedom. We remind the reader that the previous testing geometry with the same step height possessed 131,374. Therefore our trained ML model is tasked with recreating $\nu_t$ profiles from reduced observability. We observe similar results for the coefficient of friction obtained from these fine mesh assessments as shown in Figure \ref{Fig_CF_2}(b)}

\begin{figure}
    \centering
    \subfigure[Standard Mesh]{\includegraphics[width=0.48\textwidth]{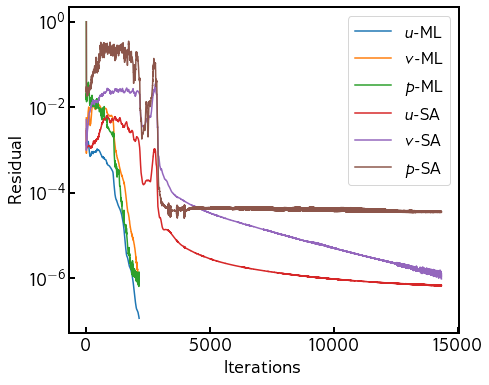}}
    \subfigure[Fine Mesh]{\includegraphics[width=0.46\textwidth]{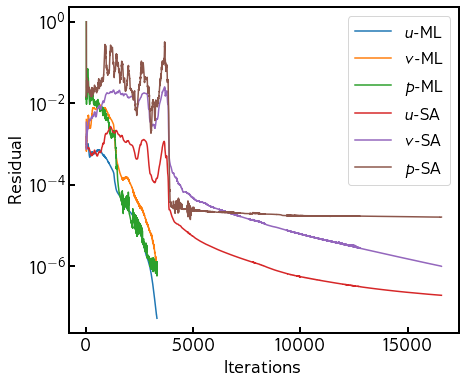}}
    \caption{Progress to convergence for the experiment assessing ML framework generalization across geometries (i.e., data-driven map $\mathbb{M}_2$). Results are shown for a step height of $1.9h$. The ML framework offers approximately 5 times faster convergence than the PDE-based model. This computational speedup is similar to the speedup for the boundary condition interpolation case in Figure \ref{Fig_Lineplot_7}. }
    \label{fig:dg_iter}
\end{figure}

\begin{figure}
    \centering
    \mbox{
    \subfigure[$\nu_t$ $L_1$-error]{\includegraphics[trim={0 6cm 0 0},clip,width=0.48\textwidth]{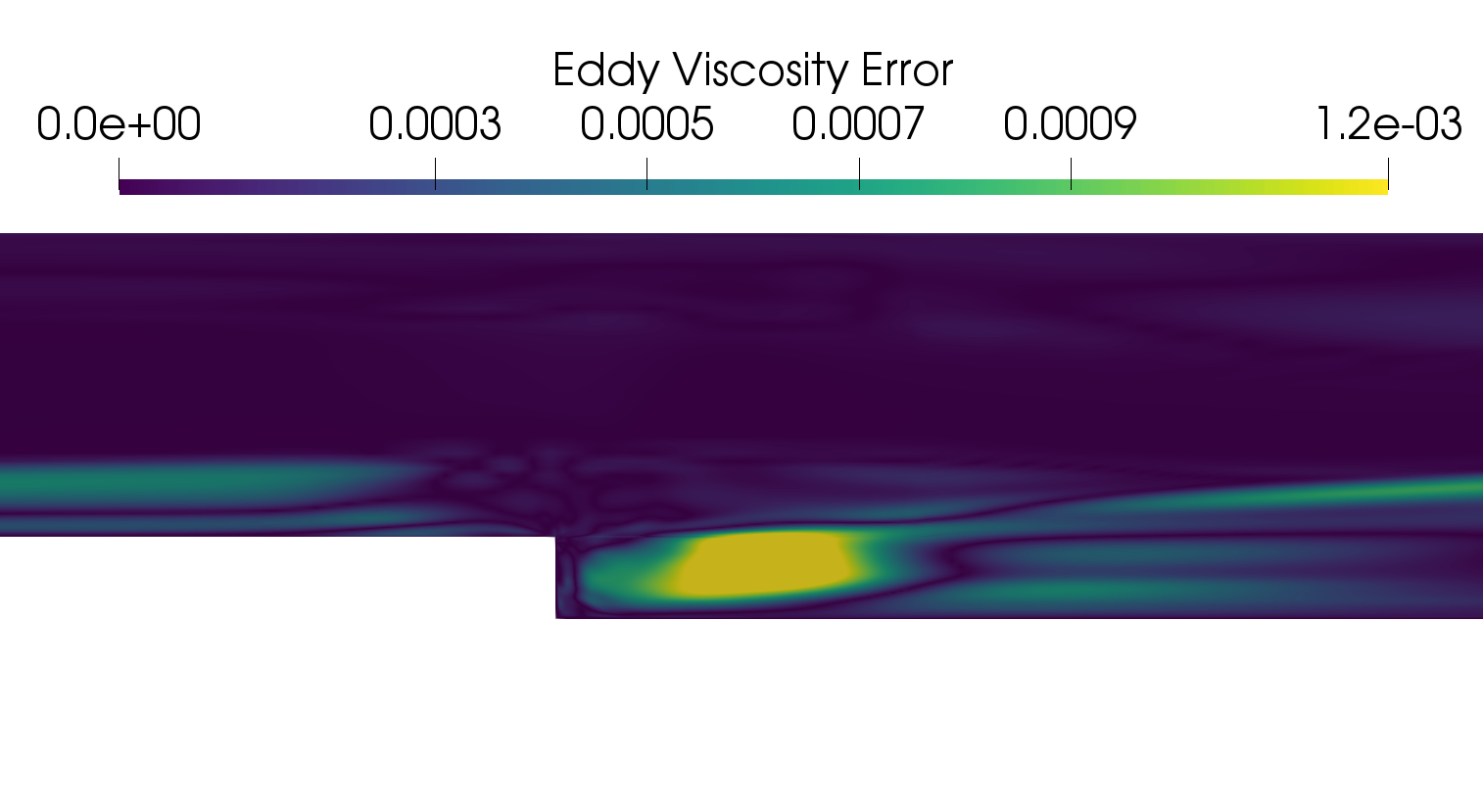}}
    \subfigure[$|U|$ $L_1$-error]{\includegraphics[trim={0 6cm 0 0},clip,width=0.48\textwidth]{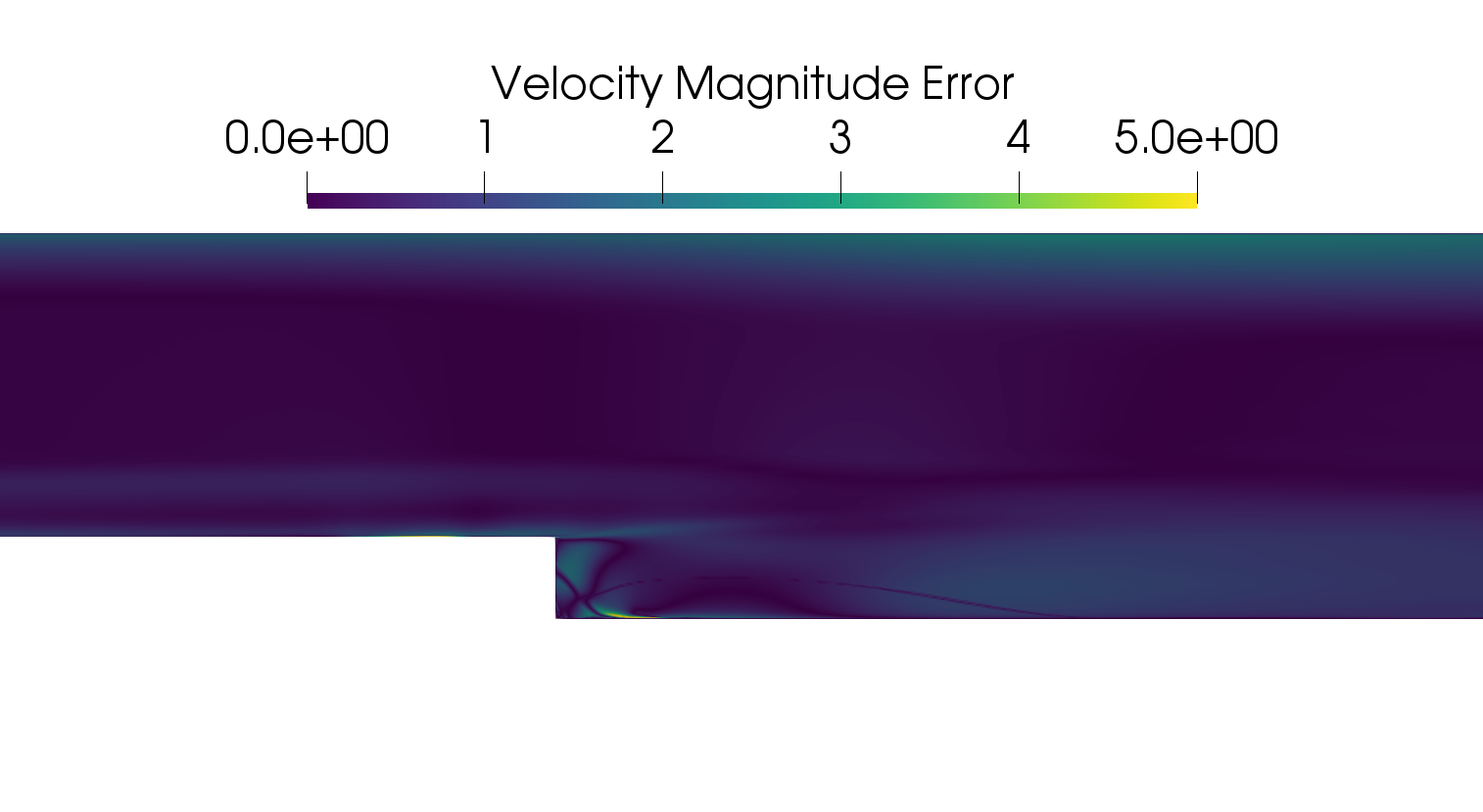}}
    }
    \caption{$L_1$-error contour plots for a backward facing step of height $1.9h$ for the data-driven map $\mathbb{M}_2$ and using a standard mesh. Note that the training set for the ML surrogate did not include this control parameter configuration. Also, errors are one order of magnitude lower than their corresponding field quantities.}
    \label{Fig_Contour_3}
\end{figure}

{Finally, the slight errors of the ML framework can be contrasted with the computational benefits of the proposed workflow which led to a converged solution in only 2132 iterations (as opposed to 14,340 iterations for the PDE-based model) for the test step height of 1.9$h$ with a standard mesh. The refined mesh required 3299 iterations (as opposed to 16,584 iterations for the PDE-based model).}. This is shown in Figure \ref{fig:dg_iter}. Contour plots for the standard mesh $1.9h$ test case are shown in Figure \ref{Fig_Contour_3} where one can observe relatively low velocity magnitude errors in regions far away from the step. The larger separation region is also recreated appropriately with the recirculation zones for step height 1.9$h$ being predicted by the proposed methodology. The circulation region shows a few hot spots where the error is relatively higher. We note that improved results are expected if a greedy-sampling of parameter space is incorporated into this workflow.

\subsection{Surrogate for two-equation models}

\begin{figure}
    \centering
    \mbox{
    \subfigure[$|U|$ $k-\omega$ SST (m/s)]{\includegraphics[width=0.5\textwidth]{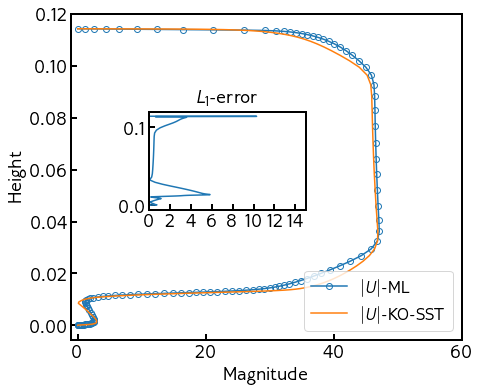}}
    \subfigure[$\nu_t$ $k-\omega$ SST ( m\textsuperscript{2}/s)]{\includegraphics[width=0.5\textwidth]{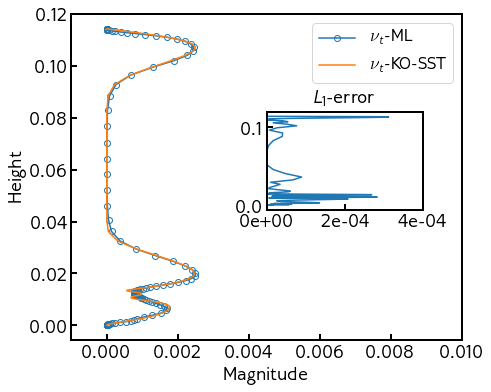}}
    } \\
    \mbox{
    \subfigure[$|U|$ $k-\epsilon$ RNG (m/s)]{\includegraphics[width=0.5\textwidth]{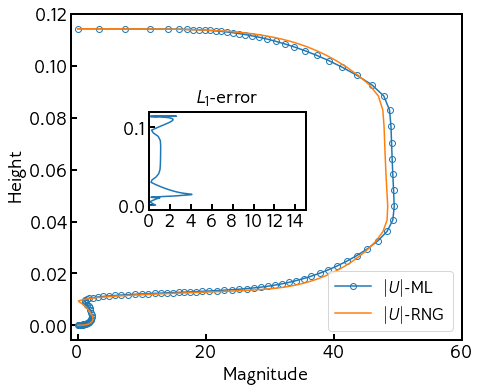}}
    \subfigure[$\nu_t$ $k-\epsilon$ RNG ( m\textsuperscript{2}/s)]{\includegraphics[width=0.5\textwidth]{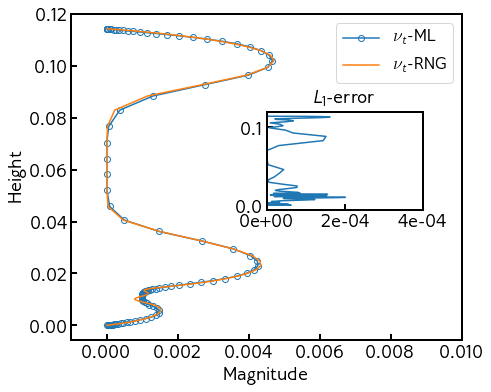}}
    }
    \caption{Surrogate modeling results for the training data generated from different closures with $k-\omega$ SST (top) and $k-\epsilon$ RNG (bottom) showing the ability of the proposed framework to reconstruct higher fidelity fields at probe location 1. This experiment learns $\mathbb{M}_1$ for two-equation eddy-viscosity models.}
    \label{Fig_KO_RNG}
\end{figure}

To demonstrate the utility of the current machine learning surrogate for turbulent eddy-viscosity models that utilize higher fidelity approximations, we also train $\mathbb{M}_1$ on training data generated from two different two-equation models, namely, the $k-\omega$ SST approximation for $\nu_t$ and the RNG $k-\epsilon$ model. We preface our results for this assessment by stating that these demonstrations are performed for training and testing solely on one boundary condition (i.e., inlet velocity) for proof-of-concept. We note that the data for this assessment was generated using first-order upwind methods whereas the SA deployments utilized second-order accurate discretizations. These discretizations were kept consistent for both the training data generation and the surrogate deployments. Figure \ref{Fig_KO_RNG} shows the results of the surrogate modeling framework for the two additional turbulence modeling strategies. {While the ML surrogate captures the near-wall trends of the eddy-viscosity profile well for both two-equation models, slight inaccuracies are observed for the transition to free-stream velocities.} We note that despite the use of lower-order methods in the training data generation, which led to a reduced number of iterations for convergence (871 and 990 iterations respectively for $k-\omega$ SST and $k-\epsilon$ RNG), the ML framework required 424 and 451 iterations respectively. This indicated a considerable acceleration as well {without compromising on accuracy}.

\section{Conclusion \& Future work}

This investigation outlines results from the development of a machine learning framework that converts the RANS closure problem to an interpolation task through the integration of machine learning. Our proposed formulation utilizes a neural network surrogate model that is trained on data generated by sampling a region in parameter space and which is deployed in the vicinity of this space for instant prediction of steady-state turbulent eddy-viscosity profiles. The inputs to the deep learning framework are given by geometry specific initial conditions leveraging the potential flow framework. The framework is tested by providing surrogates to the SA, $k-\omega$ SST, and $k-\epsilon$ RNG turbulence closures for a two-dimensional backward-facing step problem and leads to an accurate reconstruction of steady-state solutions. The workflow is also extended to a situation where different meshes (belonging to the same class of geometries) are used to train an interpolation framework spanning the range of a physical dimension given by the height of the backward-facing step. {Tests on a finer mesh for the same step height indicate that the proposed framework can generalize to some variation in mesh fidelity.} The preclusion of an additional set of equations for turbulent eddy-viscosity calculation leads to a significant reduction in the time to solution for the RANS problem using a surrogate closure. The ML surrogate trained on the SA equation to predict steady-state turbulent eddy-viscosities for an unseen initial condition is 5-8 faster than a full RANS simulation. The framework is also tested for interpolation within a range of different step heights to determine the workflow's viability for training data coming from different geometries, and similar speed-up and accuracy are obtained. A speed-up factor of approximately 2 is obtained for additional assessments on learning from two-equation models that use first-order upwinding for the advective term. We note that the formulation avoids any modification to the velocity and pressure solvers thereby leading to the preservation of the symmetries of the steady-state RANS equations. 

The conclusions from this investigation suggest the feasibility of building surrogates for sub-grid quantities for statistically steady-state problems using high-fidelity methods such as LES. Further investigation into physics regularized optimization of the steady-state mapping procedure (for instance to ensure alignment with our understanding of sub-grid mathematical properties) is also underway for greater confidence in the neural network predictions. {The current study utilizes uniformly spaced samples for the different inlet velocities or step heights for generating the training data set. The generation of a representative machine learning training data set within a limited offline computational budget is an open problem. Some promising strategies for addressing this issue include the utilization of greedy sampling strategies for sampling from a higher-dimensional parameter space. This is vital for guarding against extrapolation, where machine learning predictions deteriorate significantly. We also envision the coupling of surrogate models and training into an online feedback process where a workflow incorporates test assessments at regular intervals (or quantifies uncertainty) for the generation of a more representative data set. These costs and strategies for their mitigation will be increasingly relevant for three-dimensional simulations.} \textcolor{black}{An open area of this research is how we may be able to add interpretable outcomes from machine learning deployments, therefore, it is necessary to couple deep learning deployments with input importance assessments \cite{lundberg2017unified} for understanding the causal mechanism of the model being deployed.} The current results suggest that effective surrogate building of turbulence closure quantities using an ML framework can shift a large proportion of the online cost of parameter space exploration for assessing design quantities of interest to an offline sampling of the parameter space. \textcolor{black}{This also has applications for multifidelity frameworks where lower fidelity simulations may be generated from the proposed surrogate model to identify useful regions of parameter space for high fidelity forward model solves.}

\section*{Acknowledgments}
{We would like to thank the anonymous referee whose suggestions improved this paper considerably.}
This material is based upon work supported by the U.S. Department of Energy (DOE), Office of Science, Office of Advanced Scientific Computing Research, under Contract DE-AC02-06CH11357. This research was funded in part and used resources of the Argonne Leadership Computing Facility, which is a DOE Office of Science User Facility supported under Contract DE-AC02-06CH11357. RM acknowledges support from the Margaret Butler Fellowship at the Argonne Leadership Computing Facility. HS acknowledges support from the ALCF Director's Discretionary (DD) program for CFDML project. This paper describes objective technical results and analysis. Any subjective views or opinions that might be expressed in the paper do not necessarily represent the views of the U.S. DOE or the United States Government. Declaration of Interests - None.

\bibliographystyle{elsarticle-num-names}
\bibliography{references.bib}







\end{document}